\documentclass{aa}
\usepackage[authoryear]{natbib}
\usepackage{graphicx}

\def\smallsun{\hbox{$_\odot$}}

\def\arcsec{\hbox{$^{\prime\prime}$}}
\def\cm3{cm$^{-3}$}

\begin{document}
\voffset +0.0cm

\title{Planetary nebulae abundances and stellar evolution\thanks{Based
    on observations with ISO, an ESA project with instruments funded
    by ESA Member States (especially the PI countries: France,
    Germany, the Netherlands and the United Kingdom) and with the
    participation of ISAS and NASA.}}

\author{S.R. Pottasch\inst{1} \and J. Bernard-Salas\inst{2}  }

\offprints{pottasch@astro.rug.nl}

\institute{Kapteyn Astronomical Institute, P.O. Box 800, NL 9700 AV
  Groningen, the Netherlands \and Center for Radiophysics and Space
  Research, Cornell University, Ithaca, NY-14850-6801, USA}

\date{Received date /Accepted date}

\abstract{A summary is given of planetary nebulae abundances from ISO
  measurements. It is shown that these nebulae show abundance
  gradients (with galactocentric distance), which in the case of neon,
  argon, sulfur and oxygen (with four exceptions) are the same as HII
  regions and early type star abundance gradients. The abundance of
  these elements predicted from these gradients at the distance of the
  Sun from the center are exactly the solar abundance. Sulfur is the
  exception to this; the reason for this is discussed. The higher
  solar neon abundance is confirmed; this is discussed in terms of the
  results of helioseismology. Evidence is presented for oxygen destruction
  via ON cycling
  having occurred in the progenitors of four planetary nebulae with
  bilobal structure. These progenitor stars had a high mass, probably
  greater than 5M\smallsun. This is deduced from the high values of
  He/H and N/H found in these nebulae. Formation of nitrogen, helium
  and carbon are discussed. The high mass progenitors which showed
  oxygen destruction are shown to have probably destroyed carbon as well.
  This is probably the result of hot bottom burning.
}

\authorrunning{Pottasch \& Bernard-Salas}

\titlerunning{Planetary Nebulae Abundances}  

\maketitle

\section{Introduction}

Planetary nebulae (hereafter PNe) are an advanced stage of stellar
evolution of low and intermediate mass stars. Their abundances have
been changed by various processes which have occurred in these objects
since their formation. Not all elements have been affected in the same
way or to the same extent; some elements have not been affected in the
course of evolution. Our present purpose is to determine which
elements have been processed in the course of evolution, and which are
the affected nebulae. This is done using the abundances determined
with the help of the Infrared Space Observatory observations (ISO, Kessler 
et al.\,1996). In
particular the result of the Short and Long Wavelength Spectrometers have
been used (hereafter SWS and LWS respectively).

In order to determine which abundances have changed in the course of
evolution it is necessary to find the initial abundances before
evolution began. These values are taken from three sources: HII
regions, young (B and O type) stars, and the Sun. For the first two
groups it is known that they show a gradient in abundance as a
function of their distance from the Galactic Center (see for example
Shaver et al.\,\cite{shaver} and Rolleston et al.\,\cite{rolle}). Thus
in comparing planetary nebulae abundances with that in HII regions and
young stars this gradient must be taken into account. This is done by
plotting the abundances as function of the distance from the galactic
center and then comparing these plots.

Two potential problems arise. Firstly, the distance to individual
objects from the Sun is often uncertain. This is especially a problem
for PNe whose distances are sometimes uncertain by more
than 50\%. But because the distance to the Galactic Center depends in
an important way on the direction of the nebula from us, the distance
to the Galactic Center is actually considerably less uncertain than
the distance to the Sun.

The second problem is the following. It is assumed that the galactic
abundance gradient
arises because the material out of which the object is formed has a
different (initial) abundance at different distances from the galactic
center. The abundance in the objects measured reflects the initial
abundance at the position in which they are measured. This is easy to
defend in the case of HII regions and O and B stars because it is
unlikely that these young objects have moved far (radially) from the
position in which they have been formed.  This is not obvious for
planetary nebulae since they are older objects; they may have orbited
the galaxy one or several times since they were formed. Only if they
are on nearly circular orbits will their present position reflect the
position of formation. We will assume that this is true for the
nebulae to be considered, as is usually done (e.g. Maciel \& Costa
2003). We demonstrate in this paper that this leads to
consistent results.
 
The subject of nebular abundances has been discussed often in the past
25 years, especially from the theoretical point of view. A recent
summary has been given by Lattanzio \cite{latt} at the 2001 IAU PNe
Symposium. In ending this summary Lattanzio remarks the
following: 'What is really needed is a complete test of the model
predictions against observations, which has so far been rather
limited...'. This limitation lies in a general feeling that the theory
qualitatively explains the observations, while quantitatively little
has been done. The reason for this is, on the one hand, the uncertainty of
the theoretical results, and on the other hand the inaccuracy of the
observational abundances.

That the abundances derived from observations are often not very
accurate can be seen by a comparison of abundances derived for well
studied nebulae. Abundances derived by different authors for the same
nebula often show disagreements of a factor of 2; sometimes
considerably larger differences are found. These differences are
usually not due to differences in the observations used, and they
occur in spite of the fact that often the different authors used the
same or similar observations. The differences occur for two reasons.
The first is that the electron temperature ($T_e$) is usually assumed to
be constant in the nebula. This assumption is made because using the
optical spectrum only two ions have a sufficient number of lines to
determine $T_e$: these are \ion{[O}{iii]} and \ion{[N}{ii]}. These two
ions usually give somewhat different temperatures. The reason that one
still chooses to use a single temperature is probably that one is not
certain enough of the reality of this difference. Only after the ISO
infrared spectra became available could these temperature
variations be verified, because enough lines are now measured in many
nebulae so that electron temperatures from five to ten ions may be
obtained for a given nebula. In addition, these temperatures are
correlated with the ionization potential required to reach the
observed ionization stage. Thus a temperature gradient is often found;
the higher temperature regions occuring for the ions with the higher
ionization potentials; these probably lie closer to the center of the
nebula. From this it is now possible to measure (or predict) more
accurately the electron temperature to be assigned to each measured
ion. This in turn leads to an important improvement in the abundance
of the ion considered. Actually when the infrared lines are used,
the derived abundances are relatively insensitive to the electron
temperature since the lines are formed from very low lying energy
levels.
 
There is a second reason why the inclusion of the
infrared observations leads to increased accuracy. This is because
many more ions are now measured, and therefore there are many less
unobserved stages of ionization. This improvement in abundance
accuracy is especially important for neon, argon, nitrogen and sulfur.

One of the difficulties which are encountered in the analysis of the SWS data 
(see de Graauw et al.\,1996) occurs when the nebula is larger than 
the aperature of the instrument and a correction must be made. This is usually 
done by relating the hydrogen and helium lines in the SWS spectrum to lines of 
the same elements in other spectral regions. The theoretical line ratios then 
relate the lines in the various spectral regions. The errors introduced in this
correction are rather limited. The LWS ISO observations 
(see Clegg et al.\,1996) are made with a larger diaphragm so that no 
diaphragm correction is necessary. A more important difficulty involve the 
density dependence
of a few of the infrared lines with very low transition probabilities. In 
practice this involves the \ion{[N}{ii]} line at 121$\mu$m, the \ion{[O}{iii]}
lines at 51 and 88$\mu$m and to a lesser extent the \ion{[N}{iii]} line at
57$\mu$m and the \ion{[S}{iii]} line at 33$\mu$m. Errors can occur in low 
density nebulae if the density is not well determined or if large density 
gradients are present. For three of these ions other lines are available to
check this effect.

In this paper the abundances of 26 nebulae are considered. They are
listed in the Appendix, together with the references from which they were
taken. Their abundances have all been determined including ISO SWS and LWS
spectra and we believe are the most accurate available. They
were all determined in a similar manner, and all contain information
about electron temperature gradients in the nebulae. These nebulae
have been selected primarily because they are strong infrared
emitters. They are therefore mostly nearby PNe. They are not very much
larger than the ISO diaphragm of about 20\arcsec. Because the
constraints are mostly observational, this can be seen in a first
approximation as a random selection of nearby nebulae. For 23 of these
nebulae ultraviolet observations (IUE) are also available, without
which carbon abundances cannot be determined.
 
This is the second time that abundance results including ISO data have
been used in discussing element evolution in planetary nebulae. It was
done several years ago by Marigo et al. \cite{marigo} who used the
results of 10 PNe. In that discussion new evolution models were made
in discussing the results. We now have results from many more ISO PNe.
In addition we approach the problem of the initial abundances of the
stars which form PN in a different way. On the other hand we do not
include any self-made evolutionary models.

This paper is structured as follows. First the references are given
for the comparison objects (HII regions, early type stars and the
Sun). These are presented in Sect.\,2. In the next four sections oxygen, neon,
sulfur and argon are discussed. These are
the elements for which evolutionary abundance changes are relatively
small. In Sect.\,7 nitrogen and helium are discussed and in Sect.\,8
carbon is discussed. In each of the individual sections a comparison
is made with the abundances found in the theoretical evolutionary
models made earlier by Marigo et al. \cite{marigo}, and especially by
Karakas \cite{karakas}, which were chosen because of the large amount
of detail available in these calculations. Sect.\,9 gives a summary
and conclusions.

\section{Comparison abundances}

As discussed above, abundances from the Sun, HII regions and early
type stars will be used for comparison. In this section the references
to the sources of the data are given. The actual comparisons will be
made when discussing the individual elements. In this work it has been
assumed that the galactocentric distance of the Sun is 8 kpc. Where
necessary the distances given by individual authors have been made
compatible with this distance.

\subsection{Solar abundances}

The solar photospheric abundances are taken from the recent
compilation by Asplund et al.\,\cite{asplund} and are shown in Table 1.
All elements listed are taken from this source except for neon and
argon which have no lines in the solar photosphere. The abundance of
these two elements is taken from the measurement of coronal lines by
Feldman \& Widing\,\cite{fw}. Because the coronal lines of neon, argon
and magnesium have about the same ionization potential and strength,
the Ne/Mg(=3.6) and Ar/Mg(=0.15) ratio's are taken from this source.
The photospheric value of magnesium (taken from Apslund et al.\,2005)
is then used to find the values of neon and argon given in Table 1. Because the
solar value of neon is controvertial this will be further discussed in the 
section about neon. It should further be noted that the present solar
abundances are somewhat different than were found a decade ago.
Especially oxygen, nitrogen and carbon are now almost a factor of two
lower.

\begin{table}[t]
\begin{center}
\caption[]{ Solar Abundances. Numbers in brackets indicate: A($-$b)=
A$\times$10$^{-b}$.}
\begin{tabular}{lc}
\hline
\hline
Element &  Solar\\
        & Abundance   \\
\hline
He  &  0.085 \\
C    & 2.5(-4)  \\
N    &   0.60(-4)   \\
O   &  4.6(-4)\\
Ne  & 1.2(-4) \\
Mg   &  3.4(-5) \\
S   &  1.4(-5) \\
Cl   &  3.2(-7) \\
Ar   &  4.2(-6) \\ 
\hline 
\end{tabular}
\end{center}
\end{table}

\subsection{HII region abundances}
 
Three sources have been used for the abundances of HII regions. The
first is the work of Mart\'{i}n-Hern\'andez et al.\,\cite{martin} who have
analysed far infrared (ISO) measurements of 34 compact HII regions at
galactocentric distances between 0 and 15 kpc. They have measured
abundances of neon, sulfur, argon and nitrogen. Even though some of
the HII regions have appreciable extinction in the visible, this
problem is greatly reduced in the mid-and far-infrared. For Ne, S and Ar lines
representing all important stages of ionization are found, so that no
correction for unseen ionization stages need to be made. To obtain
abundances relative to hydrogen, the hydrogen Brackett alpha line was
used. This has the advantage over H$\beta$ both because the extinction
correction is much smaller and the diaphragm size is well defined.
Nitrogen is more difficult because the aperture size used to measure
the infrared nitrogen
lines is not the same as the hydrogen Brackett alpha line used for
comparison, as it was for neon, sulfur and argon. In the case of
nitrogen, radio continuum measurements with a similar diaphragm size
were used to obtain the abundance relative to hydrogen. In addition
only doubly ionized nitrogen is measured by ISO so that singly ionized
nitrogen must be found from optical measurements. This increases the
uncertainty of the nitrogen abundances. We have made use of their
results for HII regions with galactocentric distances from 3 to 11 pc,
which is the range for which PNe abundances are available.
 
The other sources of HII abundances are the work of Esteban et al.
\cite{esteban} and Carigi et al.\,\cite{carigi}. Esteban et al.\,\cite{esteban}
have measured the oxygen and carbon abundances for 8 HII regions
between a galactocentric distance of 6 and 11 kpc. They have used
optical spectra and have corrected for extinction using the Balmer
decrement. Oxygen abundances have been determined using both
collisional and recombination lines. Carbon abundances are only from
the recombination line at $\lambda$4267 \AA~and an ionization
correction factor (ICF) to correct for the presence of singly ionized
carbon. Carigi et al.\,(2005) have measured nitrogen abundances, but
these authors consider them to be less robust than the carbon and
oxygen. This is because only singly ionized nitrogen is measured and a
correction must be made for the more abundant doubly ionized nitrogen.

\subsection{Stellar abundances}

Abundances of main-sequence B-type stars in open clusters have been
calculated by Rolleston et al.\,\cite{rolle}. They have measured these
abundances in clusters with galactocentric distances between 5.5 and 17 kpc
(reducing the galactocentric distance of the Sun to 8 kpc). We
consider only the results below 11 kpc. The abundances of the elements
oxygen, nitrogen, carbon, magnesium, aluminum and silicon have been
derived, but we use only the first three elements for comparison.

\section{Abundance of Oxygen}

A plot of the PNe abundances is given in Fig.1. The planetary nebulae
are shown as solid squares in the figure and the HII regions as open
triangles. The dashed line is a linear fit to the PNe data (except for the
4 PNe with low O/H at 6 kpc), reproducing 
the PN points as well as possible. 
Interestingly this line also goes through the solar abundance at
R=8 kpc, 4.7x10$^{-4}$, indicating that the solar oxygen abundance is the same 
as what is
expected from PNe at this galactocentric distance. The best fit to the data 
points has a slope of -0.085 dex/kpc.

\begin{figure}[t]
  \centering
  \includegraphics[width=7.0cm,angle=90]{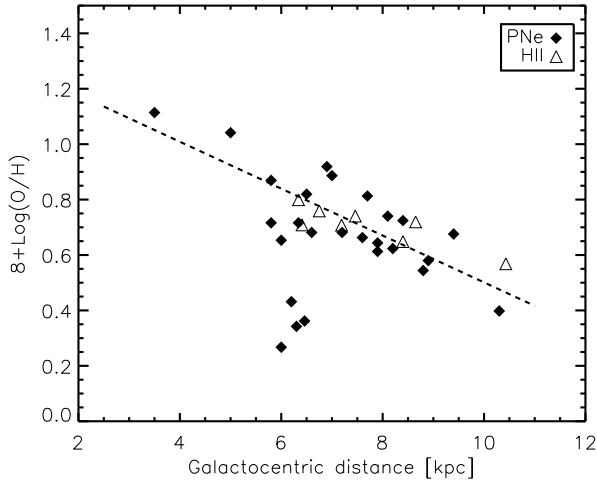}
  
  \caption{ Oxygen abundance versus the galactocentric distance.  The
    dashed line is a fit to the PNe points (see Sect.3). This line reproduces 
the solar oxygen abundance at R=8 kpc.}
        
  \label{fig-1}
\end{figure}
 
The dashed line reproduces most of the PN abundances quite well,
considering that there could be a 30\% uncertainty in the abundance
determination and an additional uncertainty in the galactocentric
distance. In addition the agreement with the HII abundances is very
good. The agreement with the B-star abundances is also good, although
on the average they seem to have a 30\% higher oxygen abundance. The B-star
abundances are not shown in Fig.1 because it makes the figure less clear. The
abundances of four PN are clearly lower than the rest by a factor of
between 2 and 3. It is suggested that the oxygen in these nebulae has
been depleted in the course of their evolution. The evidence for this
is the following.
 
First of all it is unlikely that an error of this magnitude in the
abundance determination could be made. The nebulae involved (NGC\,6302, 
NGC\,6537, Mz\,3 and He2-111) are bright and
rather easy to measure. Secondly, these nebulae have much in common.
Morphologically the first three have a classic bilobal structure;
He2-111 looks older and is irregular in shape. Another common property
is the high helium content of these nebulae, which have higher values
than any of the other nebulae whose helium has been measured. Three of the
four PNe are very high excitation objects so that although O$^{+3}$ was always
measured (and always found to be at least a factor of two less than O$^{+2}$), 
the correction for higher ionization stages is always necessary.
This was done by comparing the distribution of the ionization stages of neon,
for which every ion from Ne$^{+}$ to Ne$^{+5}$ has been measured. In addition
a photo-ionization model of NGC\,6537 was made (Surendiranath, private 
communication) which reproduces the distribution of the neon ions quite well
and confirms the lower oxygen abundance.

Thus it seems likely that for these four nebulae about half the oxygen
has been destroyed via ON cycling in the course of the evolution of the 
central stars of
these four nebulae. Marigo et al.\,\cite{marigo} have already noticed
the lower oxygen abundance of several of these nebulae but because these 
authors found it difficult
to make a model which burned sufficient oxygen and did not produce too
much nitrogen, they suggested that these stars may be descendents of
stars with lower (sub-solar) abundances. The models made by Karakas
\cite{karakas} also show some oxygen destruction in higher mass stars. In
her models with masses greater than 5M\smallsun and with Z=0.008 and Z=0.004 
Karakas also finds that too much
nitrogen is produced. But in her Z=0.02 models (solar metalicity)
with 6M\smallsun~and 6.5M\smallsun~a lower N/O
ratio is found (1.2), agreeing with observations. Because of this agreement we 
tentatively conclude
that oxygen destruction via hot bottom burning has taken place in these 
four stars and that they have masses of the order of 6M\smallsun .

\section{Neon abundance}

A plot of the neon abundances is given in Fig.2, 
The dashed line in the figure has a different meaning than the dashed
line in Fig.1. In this case it is not directly related to the points
in the figure. The dashed line is a line with the same slope as in
Fig.1 and which has (as fixed point) the solar abundance at the
galactocentric distance R=8 kpc. The fact that the points cluster
about the line means that planetary nebulae have the same neon
abundance as the Sun. That appears true of the HII regions as well.
Because of the controversy about the solar neon abundance (e.g.
Bahcall et al. 2005) we stress the point that had the solar neon
abundance of Asplund et al.\,\cite{asplund} been used, the dashed line
would lie quite far from the observed points.

\begin{figure}[t]
\centering
\includegraphics[width=7.0cm,angle=90]{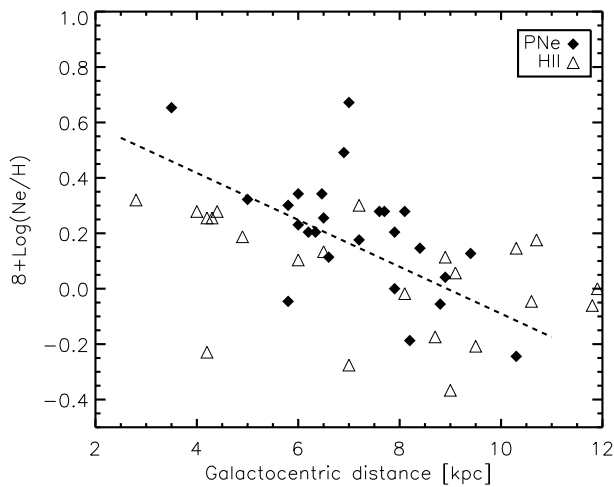}
\caption{Neon abundance as a function of the galactocentric distance.
  The dashed line is not related to the points on the diagram, but is
  a line of the same slope as the line in Fig.1 that passes through
  the solar neon abundance at R=8 kpc.}
\label{fig-2}
\end {figure}

The scatter of the individual points is somewhat larger for neon than
for oxygen, both for the PNe and for the HII regions. There is no
strong evidence, however, that neon was either created or depleted in
the course of central star evolution, to within 50\%, as there was for
oxygen. The high point at R=6.9 kpc is from NGC\,6153, which shows high
abundances for other elements as well, and is probably not an indication of
neon formation in the object. There are sometimes suggestions in the 
literature that neon
has been created in certain PNe. We have ourselves suggested that in
the case of NGC\,6302 neon has been created (Pottasch \& Beintema,
1999). We were probably misled by the rather high neon to oxygen
ratio, which is caused by the depletion of oxygen as discussed above.
In their sample Marigo et al.\,\cite{marigo} find that the neon abundances are
comparable (or slightly larger) to the solar value. To reproduce the neon 
abundance of three of their PNe (those with the largest helium and lower
oxygen abundances) under the assumption that they evolved from sub-solar
metalicity (LMC) they require a significant production of neon.
Karakas \& Lattanzio\,\cite{kl}
have suggested that some PNe may experience moderate neon enrichment. The 
latter authors show that there is only a small mass range, near 3M\smallsun,
where neon is produced in sufficient quantities to affect the neon abundance
by more than 20\%. This happens for values of Z less than 0.008. Since the 
evidence for 
enrichment greater than 50\% is lacking (as presented in Fig.2) it is possible 
that none of the PNe we have observed fall within this narrow mass range or
the neon enrichment has remained modest.

\section{Sulfur abundance}

Fig.3 is a plot of the sulfur abundances.  Since the squares and the
triangles more or less coincide, it may be concluded that the sulfur
abundance in HII regions is the same as in the planetary nebulae. The
dashed line again is a line with the same slope as in Fig.1 and which
has the solar sulfur abundance at R=8 kpc. The fact that the points
lie generally below the line indicates either that the PNe (and the HII
regions) have an abundance lower than solar or that the solar
abundance is actually higher than given by Asplund et al.\,\cite{asplund} and 
shown in
Table 1. The difference is slightly less than a factor of 2 and has 
already been noticed before by several persons and is
discussed in detail by Henry et al.\,\cite{henry} for PNe and by
Mart\'{i}n-Hern\'andez et al.\,\cite{martin} for HII regions. Henry et 
al.\,\cite{henry} suggested
that the lower sulfur abundance might be due to the incorrect ICF because they
had very few observations of the infrared [SIV] line. We can eliminate this 
possibility because this line is always observed in the ISO spectra. 

Since the PNe and solar abundances agree so well for the other elements which
are not produced in the course of evolution, it is likely that the sulfur 
abundances are actually equal. One possibility is that the solar sulfur
abundance has been overestimated by a factor of 2. While we consider this the 
most likely possibility, other possibilities exist. It is possible that the
sulfur is depleted into dust, for example. This has been suggested by
Henry et al.\,\cite{henry} but they discard it because sulfur is not 
refractory. However sulfur-based dust features (e.g.MgS and FeS) have been seen
and it would be interesting how much sulfur is contained in such features. 
Another possibility is that the atomic parameters used in
the PNe and HII analyses are incorrect. This is more unlikely because 
transitions in both the [SIII] and [SIV] lines would have to be incorrect.

\begin{figure}[t]
\centering
\includegraphics[width=7.0cm,angle=90]{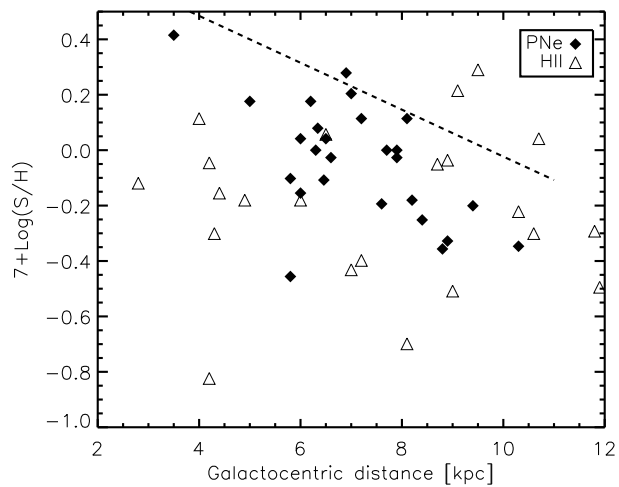}
\caption{Sulfur abundance as a function of galactocentric distance.
  The dashed line has the same slope as the line in Fig.1 while it
  passes through the solar abundance at R=8 kpc. Thus unlike oxygen
  and neon for which the PN and HII abundances agree with that of the
Sun, it appears that the solar sulfur abundance is almost twice as high as 
in PN and HII regions }
   
\label{fig-3}
\end{figure}

\section{Argon abundance}
The argon abundance is shown in Fig.4.  As can be seen from the
figure, there is good agreement between the PN and HII regions,
although there are several HII regions and one PN (Me\,2-1) which are
somewhat lower than the rest.

 \begin{figure}[t]
   \centering
   \includegraphics[width=7.0cm,angle=90]{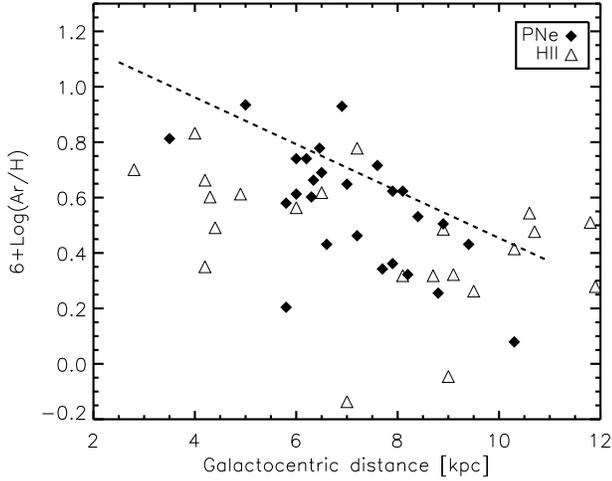}
   \caption{Argon abundance as function of the
     galactocentric distance.}
         \label{fig-4}
   \end{figure}

The dashed line is as before that line which has the same slope as that in 
Fig.\,1 and passes through the solar argon abundance at R=8\,kpc. As can be 
seen the points would be better fit if the line were slightly (between 30\% 
and 40\%) lower, indicating that a somewhat lower solar abundance would have 
been preferable. Considering the present uncertainty in the solar abundance 
(Asplund et al.\,2005) there is agreement in argon abundance in all
three types of objects.

\section{Nitrogen and Helium abundances}

%

\begin{figure}[t]
  \centering
  \includegraphics[width=7.0cm,angle=90]{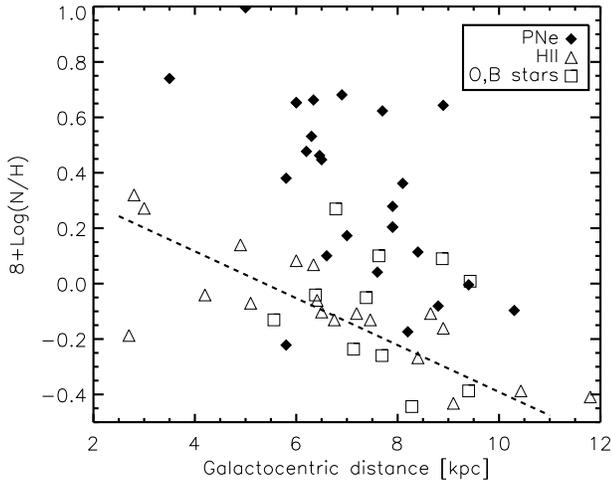}
  \caption{Nitrogen to hydrogen ratio is shown as a function of
    galactocentric distance.}
  \label{fig-5}
\end{figure}
%

The nitrogen abundance is shown in Fig.\,5.
The dashed line in the figure is, as before, not related to the
points, but is the line with the same slope as that shown in Fig.\,1
that passes through the solar nitrogen abundance at the galactocentric
distance R=8 kpc. The HII regions clearly adhere closely to the line,
as do the early type stars (with slightly more scatter). The nitrogen
abundance in planetary nebulae shows a large scatter, mostly above the
line. Only 5 PNe clearly fit to the line.  These are Me2-1 (the only
PN under the line), NGC\,7662. IC\,418, NGC\,6818 and BD+30 3639. This
is interpreted to mean that these five PNe have formed little nitrogen
in the course of their evolution. These nebulae have the common
property that their helium abundance, when known, is low i.e. about
equal to solar. Four additional PNe have a slightly higher nitrogen
abundance: NGC\,40, NGC\,2022, IC\,2165 and IC\,4191. Of these nebulae
only IC\,4191 has a somewhat higher helium abundance. Almost all the
PNe with higher nitrogen abundance show a higher helium abundance as
well.  This is shown in Fig.\,6, which is a plot of the nitrogen
abundance against the helium abundance of the sample of ISO PNe. Some
of the scatter in this diagram is because the effect of the position
in the galaxy is not taken into account. However the well known
relation between nitrogen and helium is clearly shown.  Of the five PNe
with the highest helium abundance, four are the nebulae which have
shown evidence for oxygen destruction in the course of evolution, as
discussed in section 3.

The most likely explanation of this relation is the occurance of hot
bottom burning (HBB) in higher mass stars. The models of 
Karakas\,\cite{karakas}
produce helium for high mass stars but do not find a higher value than
He/H=0.16, while at least two of our nebulae have a higher value.
Marigo et al.\,\cite{marigo} found it difficult to produce the higher helium 
values without at the same time producing more nitrogen than was observed.
They therefore chose to start with lower initial metallicity. The
Z=0.004 and Z=0.008 high mass models of Karakas\,\cite{karakas} have 
this problem as
well, but the Z=0.02 models do not overproduce nitrogen for higher
helium values. For both her 6M\smallsun~and 6.5M\smallsun~models the N/O 
ratio is about 1.2, which is similar to the observed value. The answer may 
lie in this direction.

 \begin{figure}[t]
   \centering
   \includegraphics[width=7.0cm,angle=90]{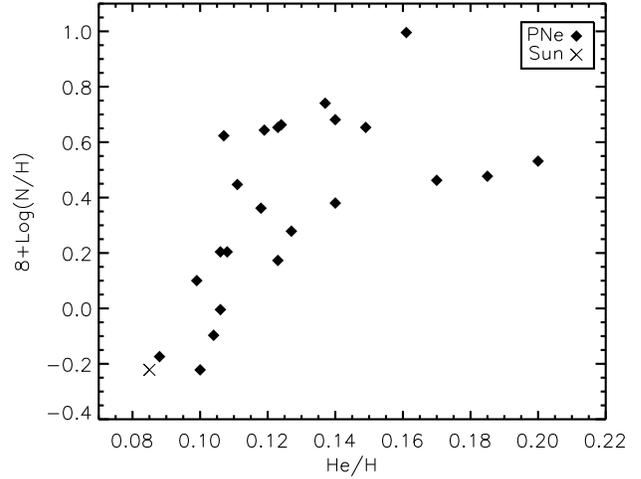}
   \caption{Nitrogen abundance plotted against the helium abundance
     for the ISO sample of planetary nebulae. }
         \label{fig-6}
   \end{figure}

\section{ Carbon abundance}

The carbon abundance as a function of galactocentric distance is shown
in Fig.7. The line has the same slope as the oxygen abundance and
passes through the solar carbon abundance at R=8 kpc. As can be seen
from the figure, the line passes quite close to the carbon abundances
of the HII regions, as it also has done for the other elements. Again note
that the HII carbon abundance is from the recombination line. The
carbon abundances of the early type stars given by Rolleston et 
al.\,\cite{rolle}
are about a factor of two less than the HII regions. Carbon is the
only element for which there appears to be a difference between the
two classes. There is some evidence that this is caused by a
systematic error in the determination of the carbon abundances by
Rolleston et al.\,\cite{rolle}. Recently Nieva \& Przybilla \cite{nieva} have 
also determined carbon abundances in six nearby early type stars and have
consistently found a higher carbon abundance. The difference is
exactly a factor of two, giving the nearby early type stars exactly
the solar abundance.

\begin{figure}[t]
  \centering
  \includegraphics[width=7.0cm,angle=90]{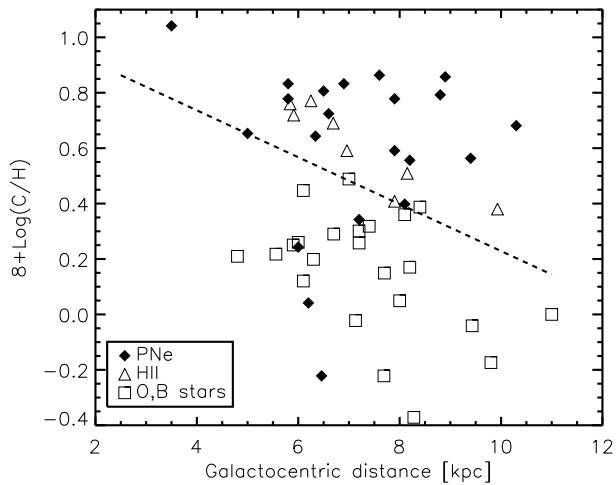}
  \caption{The carbon abundance is shown as a function of
    galactocentric distance.  The line has the same slope as in Fig.1
    and passes through the solar carbon abundance at R=8 kpc.  }
  \label{fig-7}
\end{figure}

It will be assumed that the stellar carbon abundances agree with the
line in Fig.\,7. It can be seen that the planetary nebulae carbon
abundances have quite a large scatter about this line. While about
40\% have carbon abundances agreeing with those of HII regions, about
an equal number have higher abundances, indicating that carbon has
been formed in at least 10 PNe in the course of their evolution,
presumably the effects of the third dredge-up.  This is what is
expected of stars of mass greater than about 2.5M\smallsun in the
Z=0.02 models computed by Karakas\,\cite{karakas}, and a slightly lower mass 
for the lower Z models. Marigo et al.\,\cite{marigo} predict that this will 
happen for masses of 1.5M\smallsun.

Three PNe show extremely low carbon abundances. These are the same
nebulae which have low oxygen abundances as well as high helium
abundances (there are four PNe which have these last two properties,
but no measurements of carbon lines are available in the fourth
nebula, Mz\,3). Thus it appears that carbon, as well as oxygen is
destroyed in these stars. While the higher mass models of Karakas show
initial carbon destruction due to hot bottom burning this is followed by
a few final pulses which are rich in carbon so that the net result does
not produce a decrease in carbon. The high mass models of Marigo et 
al.\,\cite{marigo} do show some carbon burning, but these models are not
the result of full evolutionary calculations.  This should be resolved by 
carefully reconsidering the models as well as reconsidering the obervations.

It is interesting to ask whether there is a relation between carbon
abundance and helium abundance as there is between nitrogen and
helium. The results are shown in Fig.\,8. The scatter in this figure
is larger than the scatter shown in Fig.\,6 (nitrogen vs. helium). A
reason for this could be that the carbon abundances are somewhat more
uncertain since the carbon lines are in the ultraviolet part of the
spectrum where the extinction plays more of a role.  The correction
for unseen ionization stages may also be more uncertain, as often only
a single ionization stage is observed. Still a clear relation is seen,
although it should be remarked that this relation depends strongly on
the three high helium nebulae.

\begin{figure}[t]
   \centering
   \includegraphics[width=7.0cm,angle=90]{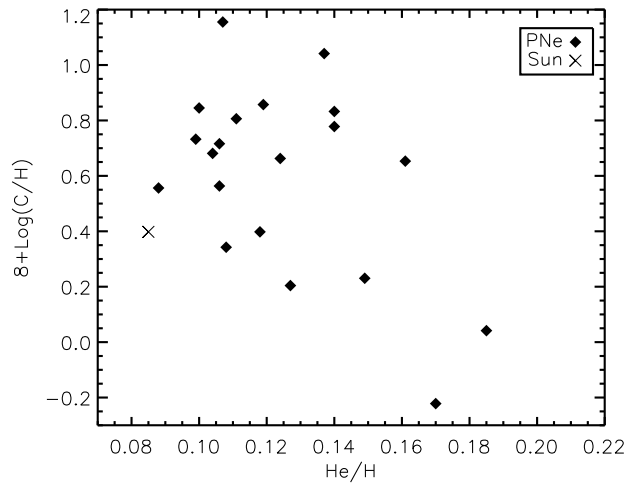}
   \caption{The carbon abundance is shown as a function of the helium
     abundance.}
         \label{fig-8}
   \end{figure}

\section{Discussion and Conclusions} 

\indent 1) For neon, sulfur, argon and most of the oxygen abundances, the
galactocentric PNe and HII values of composition are essentially
identical. This is remarkable because the HII regions are young
objects while PN are older objects. The conclusion is drawn that the
galactocentric position of the PNe does not significantly change with
time, at least for the nebulae which have been measured. This must
mean that their galactic orbits are roughly circular.  In addition all
galactocentric gradients are consistent with a single value: -0.085
dex/kpc assuming that the Sun is at 8 kpc from the Galactic Center.

2) The solar abundances of the above elements are the same as both the
PNe and HII regions at 8 kpc from the center. This is shown quite strikingly
for oxygen in Fig.\,1. Sulfur is an exception,
being almost a factor of two higher. Because it is difficult to find a
reason for this difference, the question arises as to the correctness
of either the solar sulfur abundance on the one hand or the HII and PNe
abundance on the other.

3) The solar abundance of neon is the same in the PNe and HII regions,
using the determination of the solar neon abundance given in Table 1. This
determination is very similar to that of Feldman \& Widing\,\cite{fw},
but differs from the solar neon abundance given by Asplund et al.\,
\cite{asplund}. This has significance for the controversy concerning
the consistency of solar models with helioseismological measurements
(see discussions given by Antia \& Basu \cite{antia} and Bahcall et
al.\,\cite{bahcall}). Our results support the higher neon abundance
suggested by these authors.

4) Four of the 26 PNe studied show evidence of oxygen burning in the
course of evolution. Three of these PNe have large bilobal
structures, but there are other bipolar nebulae which show no evidence
for oxygen burning. We have considered the suggestion of Marigo et
al.\,\cite{marigo} that these PNe originated from material with a lower
oxygen abundance but find the Karakas models, which are able to reproduce 
the observed N/O ratio starting with solar metalicity, a more 
likely expanation for the lower oxygen abundance.

5) The four PNe which show evidence of oxygen destruction all have high
helium abundances. Their nitrogen abundances are somewhat higher than
that of the oxygen. These four nebulae have abundances which are
compatible with model predictions of Karakas\,\cite{karakas} for progenitor 
stars with masses higher than 5M\smallsun, and a value of Z=0.02.

6) The remaining PNe show helium abundances ranging from slightly below
solar, He/H=0.088 to He/H=0.15. The nitrogen increases from about
solar to a factor of 10 above solar, directly proportional to the
helium.

7) The carbon abundance is the most difficult to interpret because it
shows the largest scatter. This is probably caused by the fact that in
some nebulae carbon is created in the course of evolution while in
others it is destroyed.  Carbon was only measured in three of the four
high mass progenitor PNe, and in all these cases the carbon was found
depleted. Of the remaining PNe about half have C/O$>$1 which is almost
certainly an indication that carbon has been formed in the progenitor
star in the course of evolution.

\begin{acknowledgements}

We thank Drs. J. Lattanzio and A. Karakas for their comments on an earlier 
version of this paper.

\end{acknowledgements}

 \appendix  
\section{Table}

The data used is given in the table, together with the references.
\begin{table*}[h]
\caption[]{Elemental abundance of PNe with ISO data in addition to optical 
and UV data.}
\begin{center}
\begin{tabular}{|l| c c c c c c c| c c | c|}
 \hline
\hline
PNe & He/H & C/H & N/H & O/H & Ne/H & S/H & Ar/H & Distance & R$^\natural$ &   Ref.,Comment\\
& &$\times$10$^{-4}$&$\times$10$^{-4}$&$\times$10$^{-4}$&$\times$10$^{-4}$&$\times$10$^{-5}$&$\times$10$^{-6}$& (kpc) & (kpc) &\\
\hline

BD+30\,3639 &         &  7.3  & 1.1  &  4.6  &  1.9  &  0.64 &  5.2  & 1.0  &  7.6  &  a   \\      
Hb\,5       &   0.130 &       & 11.5  &  6.6  &  1.8  &  1.2 &  6.3  & 3.2    & 4.8  &  b   \\
He\,2-111   &   0.185 &  1.1  & 3.0  &  2.7  &  1.6  &  1.5  &  5.5  & 2.5 & 6.2  &  c   \\
Hu\,1-2     &   0.127 &  1.6  & 1.9  &  1.6  &  0.49 &  0.42 &  1.1  & 1.5  & 7.9  &  d   \\   
IC\,418     &  $>$0.072 &  6.2  & 0.95 &  3.5  &  0.88 &  0.44 &  1.8  & 1.0  & 8.8  &  e   \\
IC\,2165    &   0.104 &  4.8  & 0.73 &  2.5  &  0.57 &  0.45 &  1.2  & 3.0  & 9.8  &  e, $\ast$ \\
IC\,4191    &   0.123 &       & 1.49 &  7.7  &  4.7  &  1.6  &  4.45 & 1.8 &  7.0  &  p   \\ 
M\,1-42    & 0.161 & 4.5  & 10.0  & 11.0  & 2.1 & 1.5 & 8.6 & 3.0 & 5.0 & h  \\ 
M\,2-36     &  0.137  & 11.0 & 5.5 & 13.0 & 4.5 & 2.6 & 6.5 & 4.5  & 3.5 & h \\
Me\,2-1     &   0.1   &  7.0  & 0.51 &  5.3  &  0.93 &  0.91 &  1.6  & 2.3 & 5.8 &  r   \\
Mz\,3       &  $>$0.080 & $<$16  & 3.0  &  2.3  &  1.2  &  1.0  &  5.0  & 2.0 & 6.3 &  q   \\ 
NGC\,40     &  $>$0.046 &  19   & 1.3  &  5.3  &  1.4  &  0.56 &  3.4  & 0.8  &  7.9  &  f   \\
NGC\,2022   &   0.106 &  3.7 & 9.9  &  4.7 &  1.3 &  0.63 &  2.7  & 1.5 & 9.4 &  p   \\
NGC\,2440   &   0.119 &  7.2  & 4.4  &  3.8  &  1.1  &  0.47 &  3.2  & 1.6  & 8.9  &  g   \\     
NGC\,5315   &   0.124 &  4.4  & 4.6  &  5.2  &  1.6  &  1.2  &  4.6  & 2.6 & 6.3  &  i   \\
NGC\,5882   &   0.108 &  2.2  & 1.6  &  4.8  &  1.5  &  1.3  &  2.9  & 1.0 & 7.2  &  e, $\ast$ \\
NGC\,6153   &   0.140 &  6.8  & 4.8  &  8.3  &  3.1  &  1.9  &  8.5  & 1.2 & 6.9  &  f   \\
NGC\,6302   &   0.170 &  0.6  & 2.9  &  2.3  &  2.2  &  0.78 &  6.0  & 1.6  & 6.4  &  j   \\
NGC\,6445   &   0.14: &  6.0  & 2.9  &  2.3  &  2.2  &  0.78 &  6.0  & 2.25  & 5.8  &  k   \\
NGC\,6537   &   0.149 &  1.7  & 4.5  &  1.8  &  1.7  &  1.1  &  4.1  & 2.0 & 6.0  &  c   \\ 
NGC\,6543   &   0.118 &  2.5  & 2.3  &  5.5  &  1.9  &  1.3  &  4.2  & 1.0 & 8.1  &  a   \\
NGC\,6741   &   0.110 &  6.4  & 2.8  &  6.6  &  1.8  &  1.1  &  4.9  & 1.65 & 6.5  &  l   \\      
NGC\,6818   &   0.099 &  5.4  & 1.26 &  4.8  &  1.27 &  0.94 &  2.7  & 1.5 & 6.6 &  p   \\
NGC\,6886   &   0.107 &  14.3 & 4.2  &  6.5  &  2.0  &  1.0  &  2.1  & 2.0  & 7.7  &  o   \\
NGC\,7027   &   0.106 &  5.2  & 1.5  &  4.1  &  1.0  &  0.94 &  2.3  & 0.65  & 7.4  &  m   \\
NGC\,7662   &   0.088 &  3.6  & 0.67 &  4.2  &  0.64 &  0.66 &  2.1  & 0.96 & 8.2  &  l   \\

\hline

\end{tabular}
\end{center}
$^\natural${Galactocentric distance assuming the Sun is at 8\,kpc from the center.}\\ 
$^\ast${Higher resolution observations.}\\ 
{{\em References:} a) Bernard-Salas et al. 2003, A\&A
406, 165, b) Pottasch et al. 2006, c) Pottasch et al. 2000, A\&A, 363,
767, d) Pottasch et al. 2003, A\&A, 401, 205, e) Pottasch et al. 2004,
A\&A, 423, 593, f) Pottasch et al. 2003, A\&A, 409, 599, g)
Bernard-Salas et al. 2002, A\&A, 387, 301, h) to be published, see also Liu 
et al. 2001, MNRAS 327, 141, i) Pottasch et al. 2002, A\&A, 393, 285, 
j) Pottasch et
al. 1999, A\&A, 347, 975, k) van Hoof et al. 2000, ApJ, 532, 384, l)
Pottasch et al. 2001, A\&A, 380, 684, m) Bernard-Salas et al. 2001,
A\&A 367, 949, o) Pottasch \& Surendiranath 2005, A\&A, 432, 139, p)
Pottasch et al. 2005, 436, 965, q) Pottasch \& Surendiranath 2005,
A\&A, 444, 861, r) Surendiranath et al. 2004, A\&A, 421, 1051.}
\end{table*}

\end{document}